\documentclass[
 reprint,
 superscriptaddress,
 showpacs,
 amsmath,amssymb,
 aps,
]{revtex4-2}
\UseRawInputEncoding
\usepackage{graphicx}% Include figure files
\usepackage{dcolumn}% Align table columns on decimal point
\usepackage{bm}% bold math
\usepackage {subfigure} 
\usepackage{hyperref}%
\hypersetup{colorlinks=true, citecolor=blue, urlcolor=blue, linkcolor=blue}

\begin{document}

\preprint{APS/123-QED}
\title{Cavity-based compact light source for extreme ultraviolet lithography}

\author{Changchao He}
\affiliation{Shanghai Institute of Applied Physics, Chinese Academy of Sciences, Shanghai 201800, China}
\affiliation{University of Chinese Academy of Sciences, Beijing 100049, China}

\author{Hanxiang Yang}%
\affiliation{Shanghai Advanced Research Institute, Chinese Academy of Sciences, Shanghai 201210, China}

\author{Nanshun Huang}
\email{nanshun@ustc.edu}
\affiliation{Zhangjiang Laboratory, Shanghai 201210, China}

\author{Bo Liu}%
\affiliation{Shanghai Advanced Research Institute, Chinese Academy of Sciences, Shanghai 201210, China}

\author{Haixiao Deng}
\email{denghx@sari.ac.cn}
\affiliation{Shanghai Advanced Research Institute, Chinese Academy of Sciences, Shanghai 201210, China}

\date{\today}%This date can be changed.
 
\begin{abstract}

A critical technology for high-volume manufacturing of nanoscale integrated circuits is a high-power extreme ultraviolet (EUV) light source. Over the past decades, laser-produced plasma (LPP) sources have been actively utilized in this field. However, current LPP light sources may provide insufficient average power to enable future manufacturing at the 3 nm node and below. In this context, accelerator-based light sources are being considered as promising tools for EUV lithography. 
This paper proposes a regenerative amplifier free-electron laser EUV source with harmonic lasing, driven by a superconducting energy-recovery linac (ERL). 
By utilizing the $n$th harmonic, the required electron beam energy is reduced to $1/\sqrt{n}$ of that in conventional schemes. The proposed configuration, employing an electron beam energy of approximately 0.33 GeV with a short-period (16 mm) undulator, is estimated to provide an average EUV power of about 2 kW.
This approach significantly reduces the required electron energy and facility size relative to other accelerator-based proposals, thereby offering new possibilities for constructing high-power EUV sources with low-energy ERLs.

\end{abstract}

\maketitle
\section{\label{sec:intro}Introduction}

The advancement of lithography technology has progressively reduced transistor sizes, significantly increasing the number of transistors on a single microelectronic chip and thereby improving computing performance \cite{fu2019euv}. Currently, chip manufacturing processes have reached the nanometer scale. Extreme ultraviolet (EUV) lithography is now the primary method for fabricating nanoscale chips, with the EUV light source being a crucial technology. EUV light sources have been extensively studied and developed over time. In current EUV lithography, a 500 W EUV light source based on laser-produced plasma (LPP) has been successfully applied to high-volume nanoscale chip manufacturing \cite{Bakshi2019EUVSF}. While LPP-based sources have been successful, they face limitations in meeting future power requirements. Insufficient EUV power leads to random pattern defects on wafers due to stochastic effects. To mitigate these effects, an EUV power exceeding 1.5 kW is required for the 3 nm node \cite{nakamura2023high}. This necessity has led researchers to explore alternative technologies, particularly accelerator-based light sources.

An accelerator-based light source is a promising tool for EUV lithography \cite{lee2020demonstration,jiang2022synchrotron,chao23storage}. Compared to LPP-based light sources, it is clean and free of debris that could contaminate mirrors in optical systems. Additionally, it can generate kW-level EUV power, and its wavelength can be easily tuned. To achieve high average power EUV, a high repetition rate electron beam is necessary, and for economic efficiency, the electron beam must have a high conversion efficiency. The energy-recovery linac (ERL) effectively meets both of these requirements \cite{hutton2023energy,hajima2010energy,bogacz2024energy}. In the ERL, a high repetition rate electron beam of more than 100 MHz can be produced using superconducting accelerator technology. After free-electron laser (FEL) emission, the electron beam is returned to the main linac for energy recovery. The low-energy electron beam from the injector and the returned electron beam alternately pass through the accelerating and decelerating RF phases in the main linac. This process transfers most of the returned energy from the electrons to the low-energy electron beam, thereby improving the utilization efficiency of the electron beam.

To produce light at a 13.5 nm wavelength with high average power based on FEL, the energy of the electron beam usually needs to exceed 600 MeV \cite{zhao2021energy,nakamura2015design,venturini2015non}. For example, KEK proposes generating approximately 10 kW of EUV light with high repetition (162.5 MHz) and 10 mA average current electron beams accelerated to 800 MeV \cite{nakamura2023high}. The main linac consists of 64 superconducting cavities and has a length of 130 m, which is too large and results in high costs \cite{kako2014development}.  Consequently, reducing the accelerator's size and lowering construction costs have become critical priorities for industrial applications of ERL-based FEL light sources. One potential approach is to utilize more advanced short-period undulators \cite{PhysRevAccelBeams.20.033201,Mishra:xe5017} to reduce the required beam energy; however, these technologies are not yet mature enough for large-scale application in short-wavelength FELs.

\begin{figure*}
\centering
\includegraphics[width=5in]{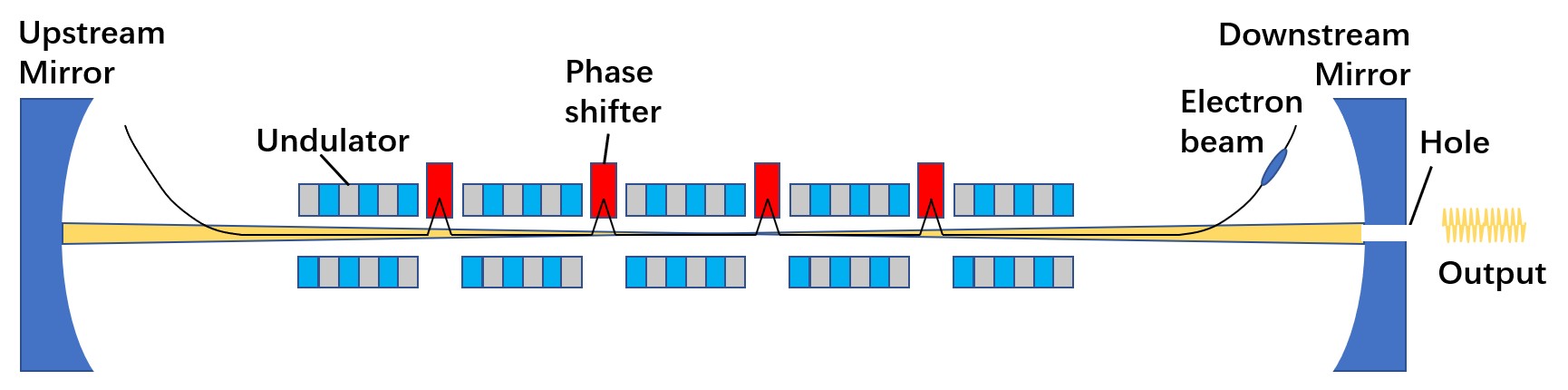}
\caption{The layout of harmonic lasing with a regenerative amplifier free-electron laser.}
\label{fig:figure1}
\end{figure*}

Here, we propose a compact EUV light source based on an ERL with harmonic lasing in a regenerative amplifier free-electron laser (RAFEL), utilizing mature techniques,  as shown in Fig.~\ref{fig:figure1}. With harmonic operation, the energy of the electron beam can be reduced to $1/\sqrt{n}$ compared to the fundamental at the same wavelength. Harmonic lasing has been studied \cite{dai2012proposal} and proven feasible \cite{neil2001second,kubarev2011third} in the cavity at other wavelengths. In RAFEL, a small fraction of the EUV radiation is reflected to interact with the electron beam in the subsequent pass, reducing the undulator length for amplification \cite{robles2023fast,tang2023active}. The RAFEL mode can also generate high pulse energy more stably compared to the SASE mode. Moreover, RAFEL can better leverage undulator tapering to further enhance power due to its improved coherence. At 13.5 nm, high-reflectivity mirrors required for RAFEL can be achieved using Mo/Si multilayer. Overall, the harmonic lasing of a RAFEL allows for lower electron beam energy, resulting in a more compact device.

\section{\label{sec:level1}Theoretical Analysis}

The resonance condition of FEL is given by \cite{huang2007review,huang2021features}:

\begin{equation}
\label{def1}
\lambda_{n} = \frac{\lambda_{u}}{2n\gamma_{r}^{2}} \left(1 + \frac{K^{2}}{2}\right)
\end{equation}
where $\lambda_{n}$ is the wavelength generated by FEL emission, $\lambda_{u}$ is the undulator period, $K$ is the dimensionless undulator strength parameter, $n$ is the harmonic number, and $\gamma_{r}$ is the electron energy in units of rest electron mass. For a fixed wavelength $\lambda_n$ and undulator parameters, the energy of the electron beam in the $n$th harmonic radiation can be reduced to $1/\sqrt{n}$ compared to the fundamental.

The proposed scheme utilizes an EUV cavity resonant at a higher harmonic of the undulator radiation, combined with phase shifting, to enable harmonic lasing of the RAFEL. This approach allows for the generation of EUV radiation at a reduced electron beam energy. In this section, we will discuss the phase shift required for harmonic lasing, analyze the efficiency of harmonic lasing, and present a possible EUV cavity design.

\subsection{ Phase Shift for Harmonic Lasing}

For constructive interaction, the phase of the electron beam and the light should match in the undulators. For example, the phase change of the electron beam is approximately $2N\pi$ between adjacent undulators, where $N$ is a integer. To enable the harmonic lasing of the FEL, the phase change could be set to $2\pi/n + 2N\pi$. This phase shift disrupts the interaction between the electron beam and the fundamental frequency while leaving the $n$th harmonic interaction unaffected \cite{mcneil2006harmonic,dai2012proposal}. Consequently, the energy of the electron beam can be more efficiently converted to the $n$th harmonic instead of the fundamental.
%after a phase shifter
\begin{figure}[h]
\centering
\includegraphics[width=3in]{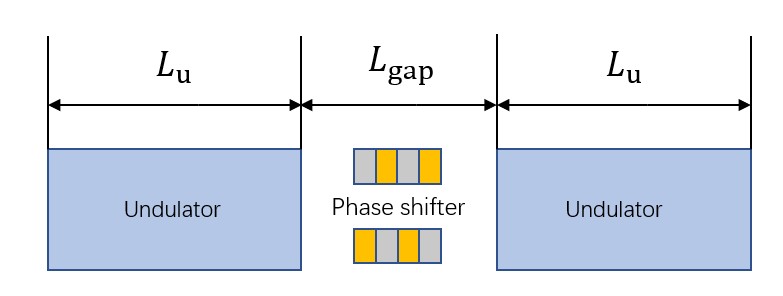}
\caption{Illustration of the gap between two adjacent undulator segments.}
\label{fig:phase_shifter}
\end{figure}

In the configuration, we assume a phase shifter is positioned between undulator segments, as illustrated in Fig.~\ref{fig:phase_shifter}. Then, the phase change  between two segments can be given by \cite{cho2023generation}:
%,freund2004phase,freund2014enhanced
%\begin{equation}
%\frac{\Delta\phi}{2\pi}=-\frac{L_{\mathrm{gap}}(1+K_{gap}^{2}/2)}
%\end{equation}

\begin{equation}
 \frac{\Delta\theta}{2\pi}=\frac{1}{2\gamma_{r}^{2}\lambda_{1}}\left(L_{\mathrm{gap}}+\left(\frac{e}{mc}\right)^{2}\cdot PI_{\mathrm{PS}}\right)
\end{equation}
\begin{equation}
PI_{\mathrm{PS}}=\int_{-\infty}^{\infty}\left(\int_{-\infty}^{z^{\prime\prime}}B_{\mathrm{y,PS}}(z^{\prime})dz^{\prime}\right)^{2}dz^{\prime\prime}
\end{equation}

where $PI_{\text{PS}}$ is the the phase integral, $B_{\mathrm{y,PS}}$
represents the magnetic field strength of the phase shifter, $L_{\text{gap}}$ is the length between two undulator segments, e is the electron charge, m is the electron mass, and c is the light velocity. The phase change includes the drift and the phase shifter.

\begin{figure}[h]
\centering
\includegraphics[width=2.5in]{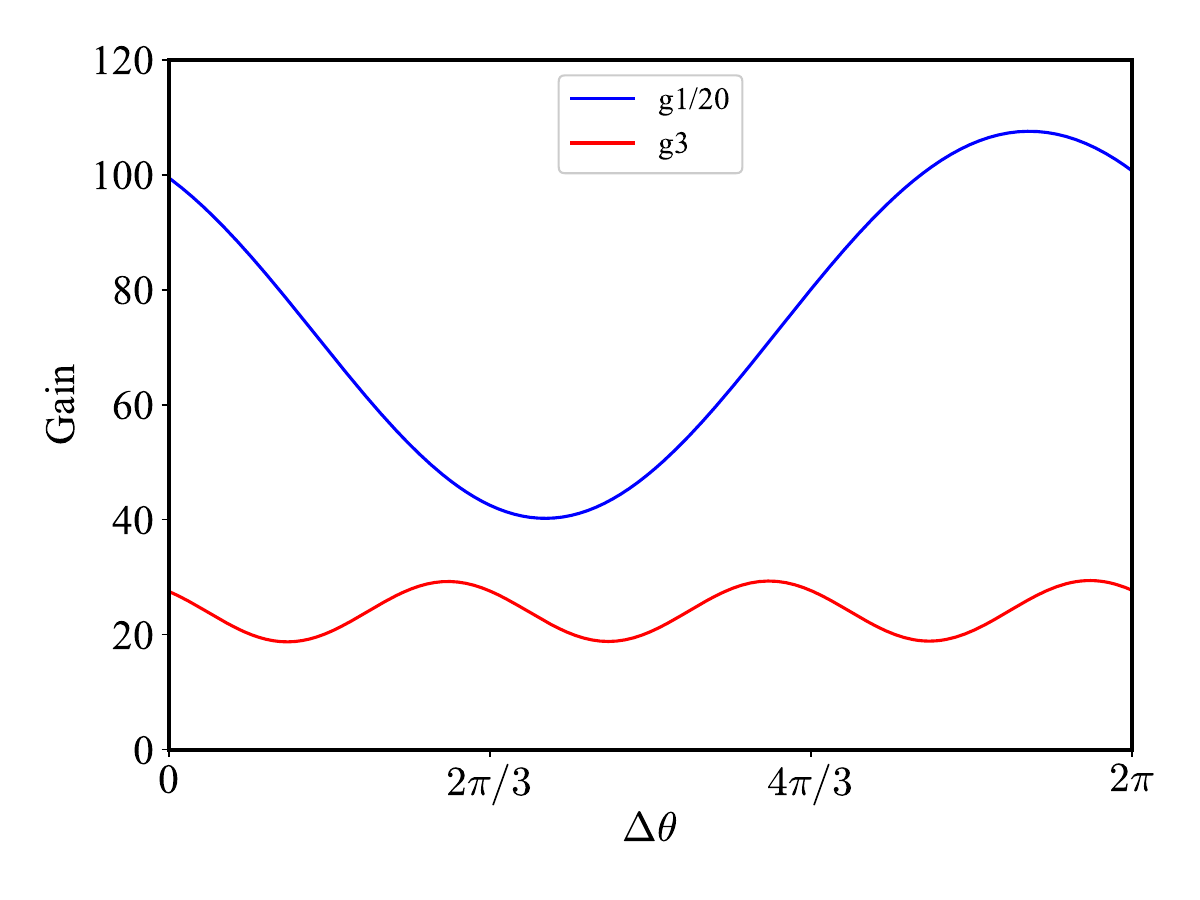}
\caption{Variation in the gain for the fundamental (blue) and the third harmonic (red) with phase jump $\Delta\theta$.}
\label{fig:gain_harm}
\end{figure}

Figure~\ref{fig:gain_harm} illustrates the numerically simulated single-pass gain of the fundamental (divided by 20) and the third harmonic, represented by blue and red lines, respectively. It can be found that phase shifter can suppress the gain of the fundamental radiation while maintaining the gain of the third harmonic. The suppression effect exhibits a periodic variation of $2\pi/n$. These results demonstrate that the FEL gain process can provide sufficient gain for RAFEL operation even with third harmonic lasing. In the subsequent simulation, the phase jump will be optimized for third harmonic amplification.

\subsection{\label{har_opt} Efficiency of Harmonic Lasing}

With the implementation of EUV mirrors, the cavity can be designed to reflect only wavelengths around the $n$th harmonic radiation. Consequently, only the $n$th harmonics can be stored by the cavity to interact with the fresh electron beam at the undulator entrance. As a result, most of the electron beam energy is converted to the harmonic rather than the fundamental radiation.

The $n$th harmonic radiation is primarily attributed to linear harmonic generation. A key scaling parameter for linear harmonics is the $n$th harmonic Pierce parameter $\rho_{n}$, defined as \cite{dattoli2005nonlinear}:

\begin{equation}
\label{def2}
\rho_{n} = \left(n \frac{[JJ]_{n}^{2}}{[JJ]_{1}^{2}}\right)^{\frac{1}{3}} \rho_{1}
\end{equation}
where
\begin{equation}
\label{def3}
[JJ]_n = (-1)^{\frac{n-1}{2}} \left[ J_{\frac{n-1}{2}}(n\xi) - J_{\frac{n+1}{2}}(n\xi)\right]
\end{equation}
\begin{equation}
\label{def4}
\xi = \frac{K^{2}}{4 + 2K^{2}}
\end{equation}
\begin{equation}
\rho_1 = \left[\frac{1}{8\pi} \frac{I}{I_{A}} \left(\frac{K[JJ]_{1}}{1 + K^{2}/2}\right)^2 \frac{\gamma_{r}\lambda_{1}^{2}}{2\pi\sigma_{x}^{2}}\right]^{1/3}
\end{equation}

Here, $[JJ]_{n}$ is the Bessel function factor for a planar undulator, which equals $[JJ]_{1}$ when $n=1$. $\rho_{1}$ represents the Pierce parameter of the fundamental. In the expression for the fundamental Pierce parameter, $I$ denotes the electron beam peak current, $I_{A} \approx 17$ kA is the Alfvén current, $\lambda_1 = n\lambda_n$ is the fundamental wavelength, and $\sigma_{x}$ is the root mean square (rms) transverse size of the electron beam.

\begin{figure}[h]
\centering
\includegraphics[width=2.5in]{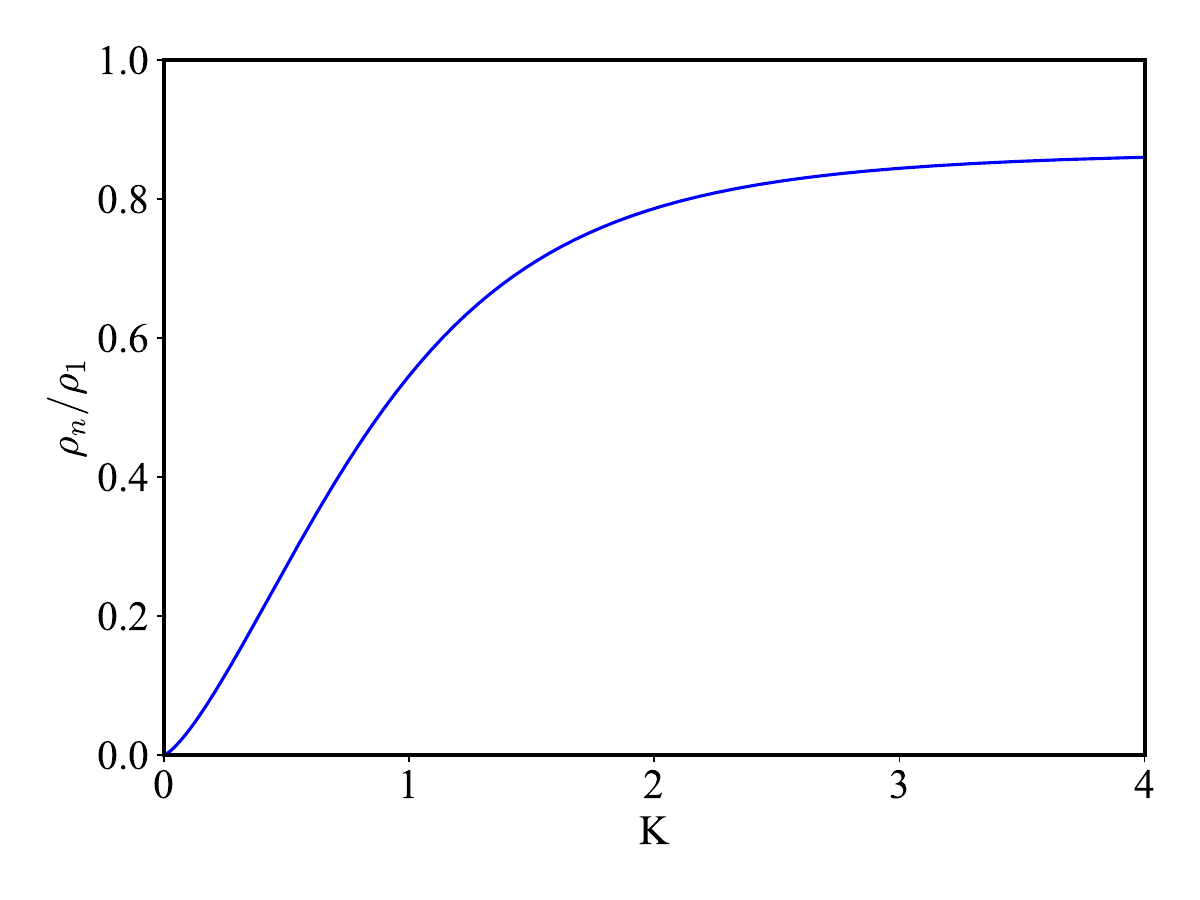}
\caption{Variation of $\rho_n/\rho_1$ with undulator strength parameter $K$ for $n=3$.}
\label{fig:K_vs_rhon}
\end{figure}

After several passes in the cavity, the pulse power of the $n$th harmonic reaches saturation, expressed as:

\begin{equation}
P_{sn} \approx \frac{\rho_{n}}{n\rho_{1}} P_{s1}
\end{equation}
where $P_{s1}$ is the saturation power of the fundamental.

As the harmonic number $n$ increases, the value of $[JJ]_n$ decreases. Therefore, $n$ should not be too large to achieve higher pulse power. In this study, we chose $n=3$. Figure~\ref{fig:K_vs_rhon} illustrates the variation of $\rho_n/\rho_1$ with the undulator strength parameter $K$ for $n=3$. As $K$ increases, $\rho_n/\rho_1$ initially rises rapidly and then gradually approaches a stable value of 0.858 for $K > 2$. However, according to the resonance relation $\gamma_{r}^{2} \propto (1 + K^{2}/2)$, an increase in $K$ necessitates a higher beam energy.

Therefore, a trade-off between the electron beam energy and the undulator $K$ value is essential, depending on the required FEL power. Here, the growth of $\rho_n/\rho_1$ begins to decelerate at approximately $K = 1.5$ in Fig.~\ref{fig:K_vs_rhon}. Hence, in this paper, we select an undulator strength parameter $K$ of 1.5 to achieve lower beam energy. This $K$ parameter can be readily achieved with an undulator period of 16~mm and a peak magnetic field of 1~T.

\subsection{EUV Opitcal Cavity}

The cavity mirrors can be composed of Mo/Si multilayer mirrors, which exhibit a reflective bandwidth of approximately 2\% at 13.5 nm. When exposed to extreme ultraviolet (EUV) light, the reflectivity of these Mo/Si multilayer films can reach up to 70\% \cite{braun2002mo}. Consequently, the cavity is designed with two Mo/Si multilayer film mirrors to ensure optimal interaction between the EUV radiation and the mirrors. \textcolor{black}{A near-hemispherical cavity design is adopted to maintain the stability of the optical field between the cavity mirrors.}  To achieve high-power extraction in RAFEL, an aperture can be made on the downstream mirror of the cavity.

The EUV light reflected through the cavity mirrors should match the time interval of the electron beam ($1/f_{\mathrm{rep}}$). The distance traveled by the EUV light in the cavity is $2L_{\mathrm{cavity}}$, where $L_{\mathrm{cavity}}$ is the length of the cavity. Therefore, without considering the slippage length, the round-trip distance of the EUV light through the cavity is:
\begin{equation}
L_{\mathrm{cavity}} = \frac{Nc}{2f_{\mathrm{rep}}}
\end{equation}
where, $c$ is the speed of light, $f_{\mathrm{rep}}$ is the repetition rate of the electron beam, and $N$ is a positive integer. Meanwhile, the cavity length $L_{\mathrm{cavity}}$ needs to be greater than the undulator length. 

%In the following simulation, the cavity length $L_{\mathrm{cavity}}$ is 20.7692 m.

\section{\label{sec:level1}Numerical Simulations and Results}

To investigate the harmonic operation of RAFEL, the Genesis code \cite{reiche1999genesis} was modified to incorporate harmonic field components into the radiation field. The propagation and coupling output of the light field in the cavity were simulated using OPC \cite{karssenberg2006modeling}. The reflection of the radiation field on the multilayer film of the cavity mirrors was calculated using the BRIGHT \cite{huang2019bright}, which is based on three-dimensional Bragg diffraction theory.

\begin{table}[h]
\caption{\label{tab:main_parameters}%
Main parameters for numerical simulation
}
\begin{ruledtabular}
\begin{tabular}{lccr}
\textrm{Parameter} & \textrm{Values} & \textrm{Unit} & \\
\colrule
Third harmonic wavelength & 13.5 & nm & \\

Beam energy & 331 & MeV & \\
Normalized emittance & 1 & mm.mrad & \\
Relative energy spread & 0.01\% & & \\

Bunch length (RMS) & 25 & $\mu$m & \\
Peak current & 370 & A & \\

Bunch charge & 77 & pC & \\
Repetition rates & 130 & MHz & \\
Average current & 10 & mA & \\

% Undulator strength $K$ & 1.5 & & \\
Undulator period & 16 & mm & \\
Undulator segment length & 2 & m & \\
Total number of undulator segments & 5 &  & \\
Inter-undulator spacing & 1 & m & \\

Reflectivity of mirror & 70\% & & \\
Cavity length & 23.077 & m & \\

\end{tabular}
\end{ruledtabular}
\end{table}

The main parameters for the numerical simulation of third harmonic lasing are presented in Table \ref{tab:main_parameters}. To generate high average power EUV radiation, a high repetition rate (130 MHz) electron beam is produced using ERL. 
The undulators have a period of 16 mm, with each segment measuring 2 m in length. A length of 1 m between adjacent undulators accommodates a phase shifter and a quadrupole magnet for electron beam focusing.
According to the FEL emission resonance condition, for a harmonic number $n=3$, undulator period $\lambda_{u}=16$ mm, and undulator strength parameter $K=1.5$, the electron beam is accelerated to 331 MeV. The average current of the electron beam in the ERL can reach 10 mA, with a bunch charge of 77 pC at a 130 MHz repetition rate. Consequently, the peak current is $\sim$ 370~A, given a 25 $\mu$m (rms) bunch length with 77 pC bunch charge. The normalized emittance is set to 1 mm.mrad, which is challenging to optimize with a 331 MeV electron beam.

In high-gain RAFEL systems,  the radiation field grows exponentially with each pass through the undulator, and the efficiency is determined primarily by the Pierce parameter. Thus, a total undulator length of 10~m is selected to slightly exceed the saturation length, to ensure sufficient gain for stable operation. In addition, the slight oversaturation helps suppress shot-to-shot energy fluctuations, contributing to better operational stability.

% In our scheme, multiple EUV radiation pulses are present in the cavity simultaneously during the propagation from the downstream cavity mirror to the upstream cavity mirror when the third harmonic reaches saturation with a 130 MHz electron beam. For simplification, in the following simulation, one pulse of radiated light is examined from initiation by noise to saturation, assuming that the fresh electron beams at the entrance of the undulators are identical.

\subsection{Undulator Tapering}

\begin{figure}[h]
\centering
\includegraphics[width=3in]{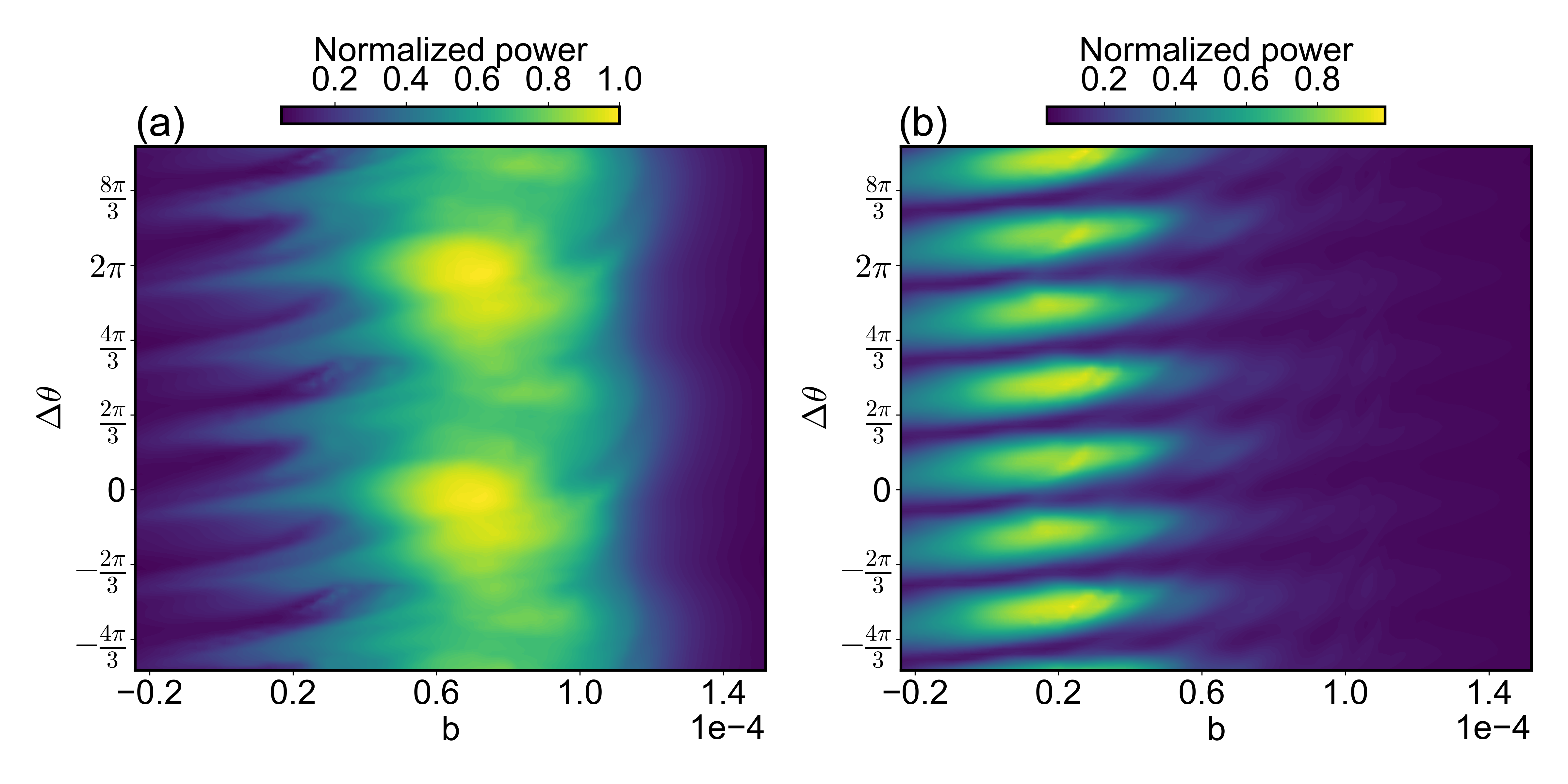}
\caption{The normalized power of (a) the fundamental and (b) the third harmonic vary with phase jump $\Delta\theta$ and taper parameter b.}
\label{fig:power_b_phase}
\end{figure}

To achieve higher output power, tapering is employed to improve the conversion efficiency from the electron beam to EUV radiation. As the electron beam amplifies radiation in the undulators, its energy decreases. Consequently, undulator tapering, which involves a reduction in the undulator parameter $K$, is necessary to satisfy the resonance condition. With undulator tapering, the formula for $K_{z}$ is given by \cite{zhao2023high}:

\begin{equation}
K_{z} = \begin{cases}
K, & \text{if} \quad z \leq z_{0} \\
K(1 - b(z  - z_{0})^{2}), & \text{otherwise}
\end{cases}
\end{equation}
Because the power of the third harmonic in the cavity increases rapidly, we set $z_{0} = 0$ m for simplification. The tapering and the phase shifter enhance the third harmonic (13.5 nm) at the same time. The normalized power of the fundamental and the third harmonic varies periodically at intervals of $2\pi$ and $2\pi/3$ respectively, shown in Fig.~\ref{fig:power_b_phase}. Meanwhile,the highest power of the fundamental and the third harmonic are located in different undulator taper. Therefore, the highest third harmonic power need phase shifter and tapering to work together. This combined approach ensures that the fundamental radiation is effectively suppressed while the third harmonic lasing is efficiently stimulated.

Additionally, for high-gain operation, the gain guiding effect must be considered while optimizing the undulator taper \cite{PhysRevLett.125.254801,li2018gain,huang2021generating,huang2024rapidly}. This consideration is crucial for RAFEL, particularly when accounting for the out-coupling of the generated EUV light through a hole in the downstream mirror. The hole should be optimized for a stable transverse mode, which must match the gain guiding for high output coupling. Consequently, simultaneous optimization of the hole radius and undulator taper is necessary to achieve a stable transverse mode and high output power.

\begin{figure}[h] 
    \centering
    \includegraphics[width=3in]{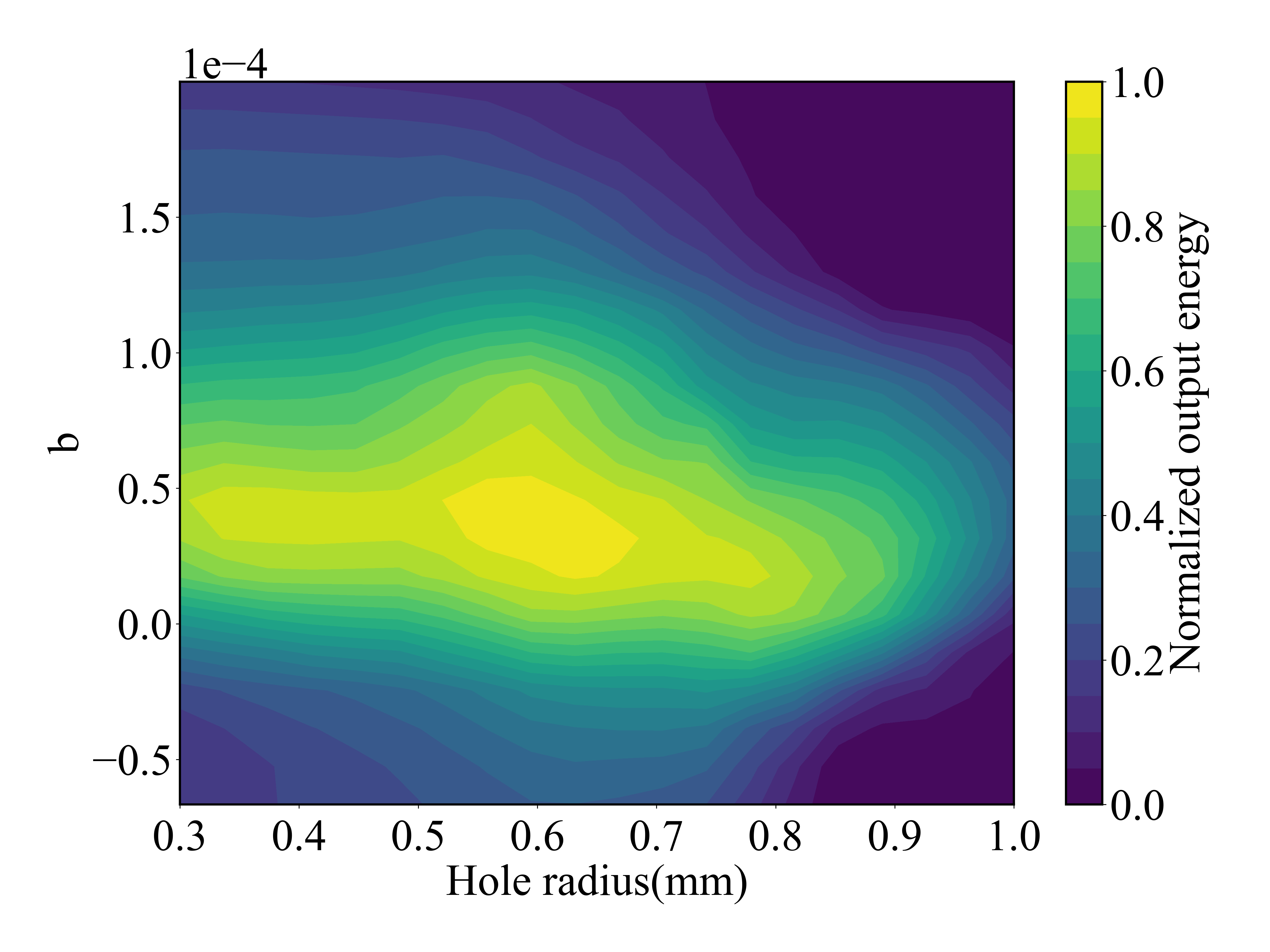}
    \caption{Normalized output pulse energy with different hole radius sizes and different taper parameters $ b $.}
    \label{fig:figure5}
\end{figure}

Figure~\ref{fig:figure5} illustrates the normalized output pulse energy for various hole radius and taper parameters $b$. To achieve high extraction efficiency, RAFEL tends to utilize a larger hole. However, it is evident that with a large hole size, only a narrow range of undulator taper values can match the transverse mode to produce high power. The optimal hole radius spans a relatively wide range around 600 $\mu$m, with the optimal taper parameter $b$ ranging from $0.2 \times 10^{-4}$ to $0.5 \times 10^{-4}$. Accounting for potential manufacturing process errors, we selected a hole radius of 650 $\mu$m and a taper parameter $b$ of $0.3 \times 10^{-4}$. The out-coupling rate is around 72\%.

\subsection{Cavity Detuning}

The cavity detuning in RAFEL is distinct from that observed in a low-gain oscillator. 
In a low-gain oscillator, the slippage length is defined as $l_{\mathrm{slip}} = N_{u}\lambda_{1}$, whereas in a high-gain RAFEL, it is given by $l_{\mathrm{slip}} = N_{u}\lambda_{1}/3$~\cite{freund2013three}. Here, for simplification $N_u$ is the number of periods in the undulator length including the phase shifter. This difference is attributed to the exponential growth, which reduces both the phase and group velocities.
% In these expressions, $N_{u}$ represents the number of undulator periods, as detailed in Table~\ref{tab:main_parameters} in the simulation, and $\lambda_{1}$ corresponds to the fundamental wavelength. 
Then, the cavity detuning can be defined as $\Delta L = L_\mathrm{cavity} - L_0$. The expected optimal cavity detuning in a high-gain RAFEL is around the negative value of the slippage length, $-N_{u}\lambda_{1}/3$. 

\begin{figure}[h]
\centering
%\includegraphics[width=1.6in]{figure6}
% \caption{The cavity detuning curve and the corresponding jitter of the third harmonic output pulse energy.}
\includegraphics[width=3.5in]{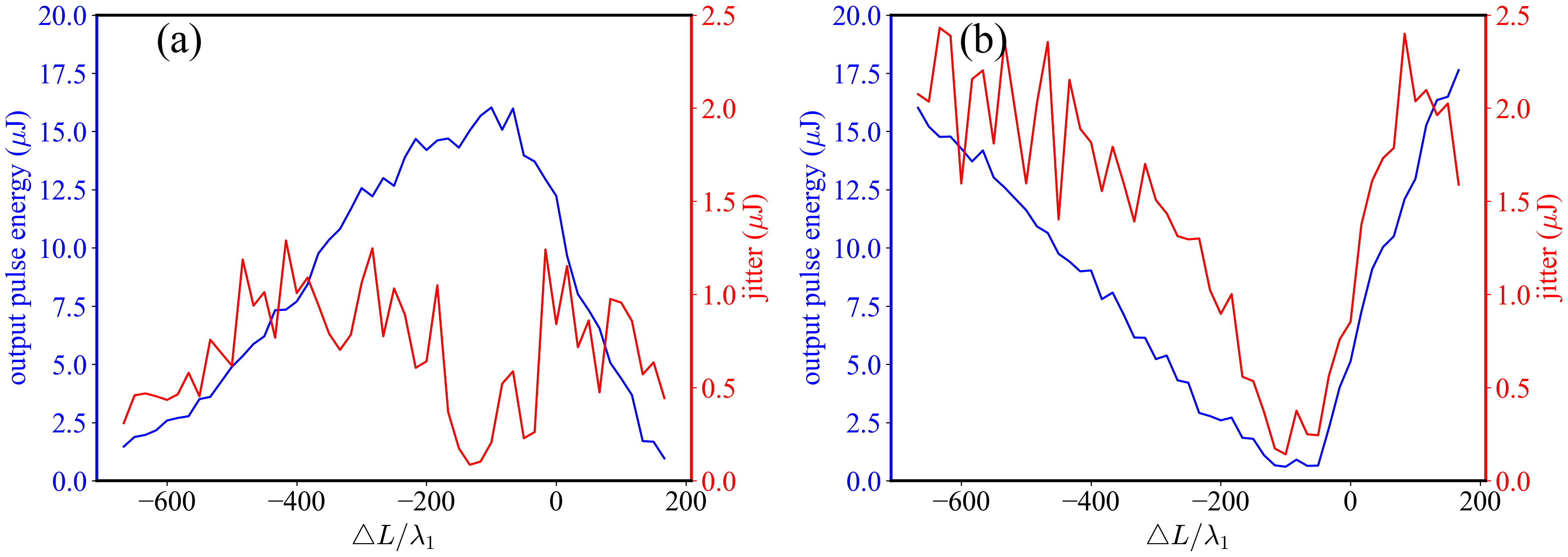}
\caption{The cavity detuning curve and the corresponding jitter of (a) the third harmonic and (b) the fundamental output pulse energy.}
\label{fig:figure8}
\end{figure}

Figure~\ref{fig:figure8}(a) illustrates the cavity detuning curve. 
At the peak of output pulse energy, the optimal cavity detuning is determined to be in range from $-300$ to $-100 \lambda_{1}$, which closely approximates the value of $-N_{u}\lambda_{1}/3$. The variation in the rms output pulse energy jitter is also presented. When At the peak of output pulse energy, the relative pulse energy jitter is approximately 5\%.

% \begin{figure}[h]
% \centering
% \includegraphics[width=2.5in]{figure8}
% \caption{The variation in the standard deviation of output pulse energy from pass 15 to pass 50.}
% \label{fig:figure8}
% \end{figure}

Compared to the third harmonic, fundamental radiation show different trends in output pulse energy and jitter as a function of detuning. As shown in Fig.~\ref{fig:figure8}(b), the fundamental radiation can be reduced to 0.6 $\mu$J per pulse, which is more than an order of magnitude lower than the energy of the third harmonic. While there is a small amount of fundamental radiation present at the source, the optical system inherent filtering characteristics ensure that it does not affect the intended operation, as the Mo/Si multilayer mirrors used in the EUV light transport system have a very low reflectivity at $\lambda_{1} = 40.5$ nm.

% It can be found that larger pulse energies correspond to greater jitter in pulse energy. When the cavity detuning is close to $-N_{u}\lambda_{1}/3$, the relative pulse energy jitter is approximately 5\%. When the cavity detuning significantly deviates from $-N_{u}\lambda_{1}/3$, the pulse energy jitter is substantially reduced.
% This reduction is attributed to the decreased interaction between the EUV light reflected through the cavity mirrors and the electron beam, leading to diminished output pulse energy and energy jitter.

\begin{figure}[h]
\centering
\includegraphics[width=2.5in]{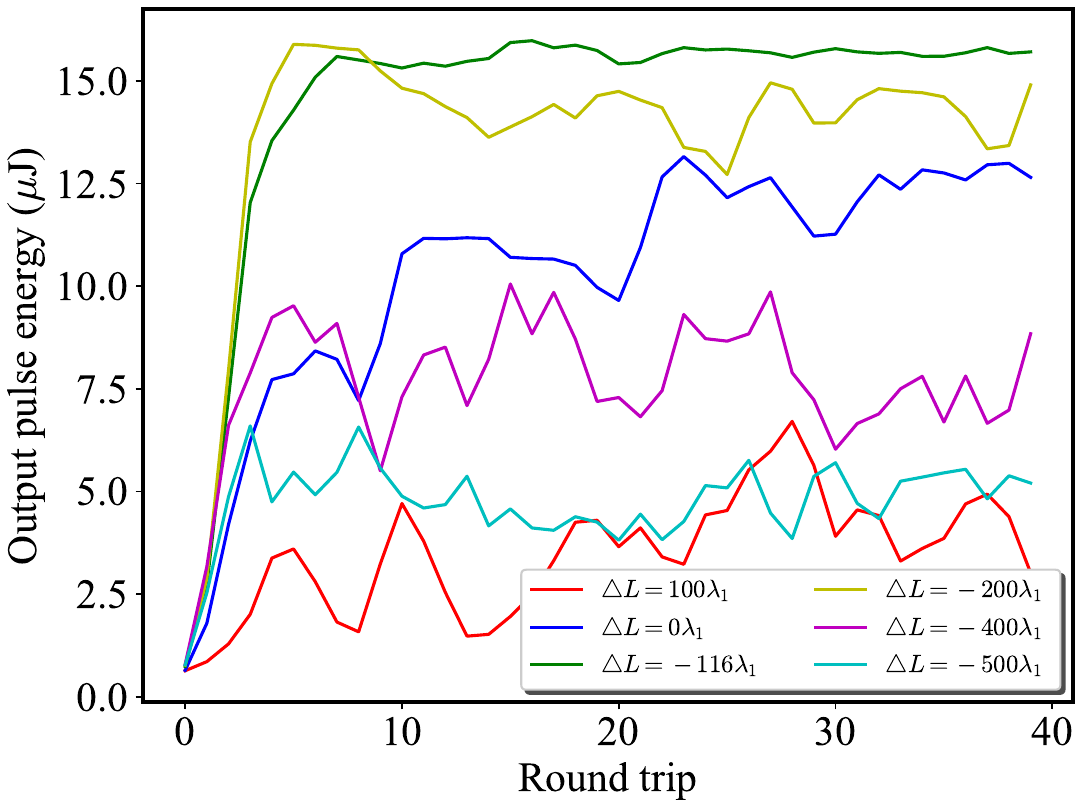}
\caption{Output pulse energy evolution of the third harmonic for various cavity detuning.}
\label{fig:figure7}
\end{figure}

% Similar energy jitter, referred to as limit-cycle oscillations, have been observed in low-gain oscillators and in simulations of RAFEL \cite{jaroszynski1993limit,freund2013three}. 
% One possible reason for these fluctuations is the high gain and small fraction of EUV light returning to the undulator in RAFEL, which means the output EUV light is averaged by few pass of FEL amplification.
% Additionally, the spectral characteristics and stability of RAFEL are intermediate between those of a low-gain oscillator and a Self-Amplified Spontaneous Emission (SASE) mode. 
% Consequently, minor changes in the EUV spectrum and pulse energy significantly affect the interaction between the electron beam and the reflected EUV light. 
% Besides, the reflectivity of a cavity mirror is only 70\% and a bandwidth of only 2\%, thereby increasing the instability of EUV light.
% The EUV mirror has a reflection bandwidth of only 2\%, thereby increasing the instability of EUV light. 

Figure~\ref{fig:figure7} depicts the detailed energy evolution of the third harmonic with each round trip, where cavity detunings range from $-600\lambda_{1}$ to $200\lambda_{1}$.
% \textcolor{red}{It can be found that larger pulse energies correspond to greater jitter.} 
The energy jitter generally fluctuates but exhibits a decreasing trend with increasing detuning until $\Delta L \simeq -200 \lambda_1$, after which it begins to rise.
Similar energy jitter, referred to as limit-cycle oscillations, has been observed in low-gain oscillators and in simulations of RAFEL \cite{jaroszynski1993limit,freund2013three}. One possible explanation for these oscillations is the combination of high gain and the small fraction of EUV light returning to the undulator in RAFEL. The changes in EUV spectrum and pulse energy can significantly affect the interaction between the electron beam and the reflected EUV light. This sensitivity is further exacerbated by the properties of the cavity mirrors. The reflectivity of a cavity mirror is limited to 70\%, which reduces the amount of light available for subsequent amplification passes. Moreover, the EUV mirror has a narrow reflection bandwidth of only 2\%, which further contributes to the instability of the EUV light.

%

% One possible explanation for these oscillations is the combination of high gain and the small fraction of EUV light returning to the undulator in RAFEL. This configuration results in the output EUV light being averaged over only a few passes of FEL amplification. Consequently, the system becomes more susceptible to fluctuations in the amplification process.
% Thus, changes in the EUV spectrum and pulse energy can significantly affect the interaction between the electron beam and the reflected EUV light. This sensitivity is further exacerbated by the properties of the cavity mirrors. The reflectivity of a cavity mirror is limited to 70\%, which reduces the amount of light available for subsequent amplification passes. Moreover, the EUV mirror has a narrow reflection bandwidth of only 2\%, which further contributes to the instability of the EUV light.

From the LPP source standard, it requires dose stability of approximately 0.5\% during single exposure. Using a 4-millimeter slit height at the photomask with a scanning speed of 1 meter per second, each exposure takes 4 milliseconds \cite{anderson2024compatibility,mizoguchi2023plasma}. At the operational pulse rate of 130 MHz, this translates to 520,000 pulses per exposure. This high number of pulses provides significant statistical averaging, reducing energy jitter by a factor of $\sqrt{520,000} \simeq 721$. Consequently, the RAFEL output would meet the required dose stability specifications even with large jitter of 20\%.

% Furthermore, the high hole output coupling ratio (over 70\%) affects the transverse mode structure of the seeding light.

\subsection{Output Performance and Characteristics}

\begin{figure}[h]
\centering
\includegraphics[width=3.5in]{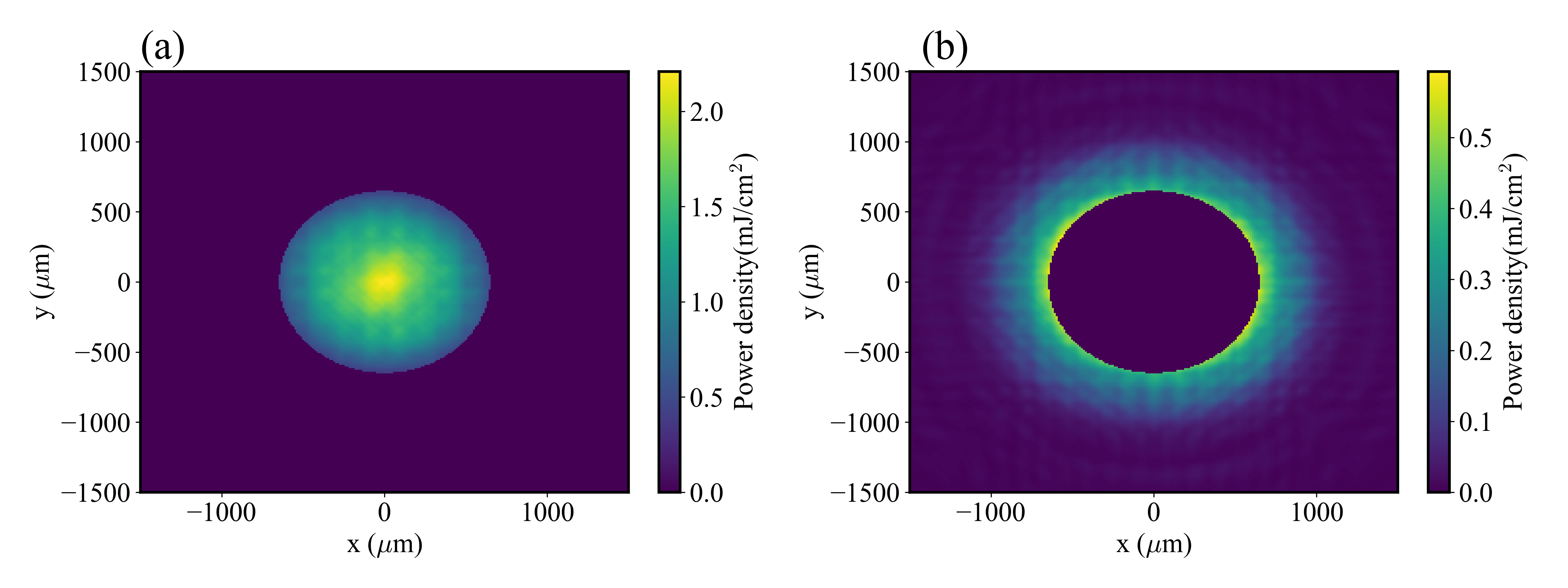}
\caption{The third harmonic power density of (a) output pulse and (b) pulse on the downstream mirror.}
\label{fig:power_density}
\end{figure}

% By allowing the reflected EUV radiation to recirculate in a 23.077 m round-trip cavity and interact with the electron bunches multiple times, the maximum energy at undulator exit is increased to 21.5 $\mu$J. 
Steady-state RAFEL behavior is achieved in approximately six passes. The out-coupling rate is determined to be about 72\%. An average saturated output energy of $\sim$15.6 $\mu$J is transmitted through the cavity hole, resulting in an average output power exceeding 2 kW for the 130 MHz electron beams. The energy density per pulse of the EUV RAFEL light at normal incidence is depicted in Fig.~\ref{fig:power_density}. A maximum power density of approximately 2.3 mJ/cm$^2$ is observed with a transverse size of $ \sim$0.3 mm at the exit, while the power density at the Mo/Si mirror is only 0.6 mJ/cm$^2$. These values are significantly lower than the ablation thresholds of Mo/Si multilayer and Si, which have been experimentally estimated to be about 20 mJ/cm$^2$ by SACLA-BL1 \cite{NishikinoEUV} and FLASH \cite{makhotkin2018experimental}.

In general, the RAFEL outputs are transversely coherent. Thus, a newly illumination system in Ref.~\cite{pistor2024exploring,anderson2024compatibility} is designed to effectively manage transverse coherence, which transversely splitting each FEL pulse into approximately 400 individual pulses. The temporal spread of these pulses ranges from 10 to 100 picoseconds, resulting from millimeter-scale path differences between illumination channels. Thus, while individual FEL pulses are transversely coherent, the integrated exposure illumination is effectively incoherent, as the illumination system serves as a coherence reducer.

Besides, in the proposed setup, the maximum heat load on the downstream cavity mirrors is estimated to be approximately 350 W (with fundamental and third harmonic power). Thus, liquid nitrogen cooling should be employed to maintain the mirrors within a safe operating temperature range. Besides, to enhance thermal contact and improve heat transfer efficiency, an indium foil should be applied between the heat sink and the mirror substrate. To further reduce thermal stress, it can lower the power density on downstream mirrors by modifying the distance between the downstream mirror and the undulator. 

\begin{figure}[h]
\centering
\includegraphics[width=3.5in]{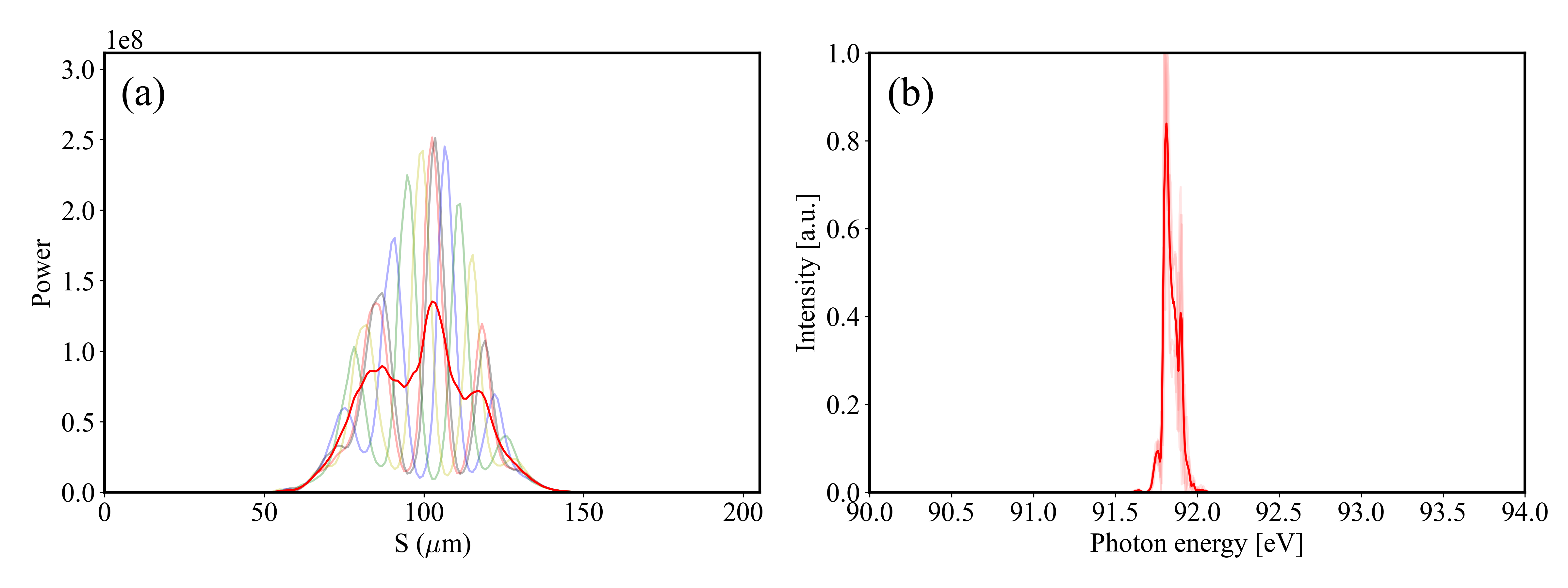}
\caption{(a) The temporal power profile and (b) spectrum of output pulse when reaching saturation.}
\label{fig:power_profile_spectrum}
\end{figure}

The temporal profile and spectrum of the 13.5 nm output pulse at saturation are presented in Fig.~\ref{fig:power_profile_spectrum}. The EUV output is characterized by a spiky temporal profile and spectrum, with power reaching up to 250 MW. The FEL spectral width is less than 0.3 eV, which is sufficiently narrow for the reflectivity of the following Mo/Si mirror. Because, compared to the LPP sources, this narrower spectrum aligns more efficiently with the multilayer mirror reflectivity curve, resulting in reduced energy loss through absorption at each mirror interaction. Therefore, the RAFEL inherently narrower spectrum provides a significant advantage~\cite{pistor2024exploring}.

% This spiky nature differs from that observed in crystal cavity-based RAFELs, where the spectrum typically exhibits a smooth single spike due to a relative bandwidth of only ~1e$^{-5}$ \cite{tang2023active}. The observed spikiness is similar to that predicted for long wavelengths with a large reflective bandwidth mirror. This similarity can be attributed to the EUV multilayer mirrors, which possess a high reflective bandwidth of 2\%, exceeding that of typical SASE output. Additionally, the spiky spectrum is a consequence of pulse oversaturation, resulting in properties comparable to, or potentially worse than, the equivalent SASE case \cite{DUNNING2008116}. 

Overall, the harmonic lasing RAFEL can significantly reduce the required electron beam energy without notably decreasing the FEL performance. With harmonic lasing, the beam energy is significantly decreased from 800 MeV to 330 MeV, compared to the high-power EUV FEL based on the KEK design \cite{nakamura2023high}. This reduction in energy allows for a potential decrease in footprint size from $200 \times 20$~m$^2$ to $80 \times 10$~m$^2$. Furthermore, the proposed harmonic lasing RAFEL is compatible with advanced compact multiturn ERLs, which could further reduce the footprint size \cite{socol2011compact}. 

Additionally, the industry trend toward shorter wavelengths for smaller feature sizes, particularly using the 6.x nm region (Blue-X) \cite{bakshi2023photon}. The flexibility of FEL scheme allows for wavelength adjustment, which is a distinct advantage over LPP sources. Recent advances in mirror technology, achieving 60\% reflectivity at 6.x nm \cite{uzoma2021multilayer}, make the proposed harmonic RAFEL operation viable at these shorter wavelengths. 
Ultimately, the proposed scheme significantly reduces the required electron energy and facility size, offering new possibilities for constructing low-energy ERLs to test relevant technologies and for future industrial applications in EUV lithography.

\section{Conclusion \label{sec:concl}}

In conclusion, the integration of harmonic lasing and RAFEL with an ERL has been demonstrated to facilitate the development of a compact EUV light source suitable for industrial applications.  The utilization of the $n$th harmonic allows for a reduction in the electron beam energy to $1/\sqrt{n}$ of the conventional scheme. Specifically, 13.5 nm EUV light with an average power exceeding 2 kW can be produced by a RAFEL operating at the third harmonic, driven by an ERL with an electron energy of merely 0.33~GeV and average current of 10~mA. The peak power reaches up to 250 MW, while the spectral width remains below 0.3 eV, which is sufficiently narrow to maintain the reflectivity of the Mo/Si mirror. To further minimize the facility size, higher harmonic numbers (e.g., $n=5$) may be considered. Additionally, the implementation of a multiturn ERL scheme could further contribute to reducing the overall facility size.

This method can be extended to shorter wavelengths. The incorporation of a crystal cavity could enable the production of X-ray radiation. For instance, the utilization of 4~GeV electron beams in conjunction with an X-ray RAFEL could potentially facilitate the generation of 10 keV X-rays characterized by narrow bandwidth and high peak power.

\begin{acknowledgments}
This work was supported by the National Natural Science Foundation of China (12241501, 12125508), the National Key Research and Development Program of China (2024YFA1612100), the CAS Project for Young Scientists in Basic Research (YSBR-042), and Shanghai Pilot Program for Basic Research – Chinese Academy of Sciences, Shanghai Branch (JCYJ-SHFY-2021-010), Natural Science Foundation of Shanghai (22ZR1470200),the China Postdoctoral Science Foundation (2023M733014, 2024T170758).
\end{acknowledgments}

% \bibliography{apssamp}
\providecommand{\noopsort}[1]{}\providecommand{\singleletter}[1]{#1}%

\end{document}